\documentclass[aps,prl,showpacs,twocolumn]{revtex4}
\newcommand\llangle{\langle\!\langle}
\newcommand\rrangle{\rangle\!\rangle}

\usepackage{graphicx}
\begin{document}

\title
{
\begin{minipage}[t]{7.0in}
\scriptsize
\begin{quote}
\leftline{{\it Phys. Rev. Lett.}, in press.}
\raggedleft {\rm arXiv:0907.5512}
\end{quote}
\end{minipage}
\medskip

New Correlated Model of Colossal Magnetoresistive Manganese Oxides}

\author{D. I. Golosov}
\email{golosov@phys.huji.ac.il}
\affiliation{Department of Physics and the Resnick Institute, Bar-Ilan 
University, Ramat-Gan 52900, Israel.}

\date{\today}

\begin{abstract}
A new minimal model is constructed for the doped manganese oxides which exhibit
colossal magnetoresistance (CMR), involving a broad spin-majority
conduction band as well as nearly localised spin-minority
electron states. A simple mean field analysis yields a temperature-dependent
hybridised band structure with suppressed carrier weight at the Fermi level.
Spin stiffness is complex, indicating strong spin wave damping.
Further investigations are needed 
to  verify the relevance of the proposed model.

\typeout{polish abstract}
\end{abstract}
\maketitle

The unusual properties of doped manganese oxides exhibiting 
colossal magnetoresistance (CMR) \cite{Tokurabook} are not yet
understood theoretically, and the problem of formulating
a suitable microscopic model remains open. These properties 
include the CMR phenomenon and metal-insulator transition, 
which are, in turn, intimately related to the temperature-induced 
variation
of the electron density of states (``pseudogap'' \cite{pseudogap}) or of the 
effective
carrier number\cite{Okimoto}.
This implies that, alongside double exchange ferromagnetism, 
the effects of electron-electron interaction play a key
role 
and, quite possibly,
are responsible for CMR itself. Indeed, the bandstructure
calculations \cite{bandcoulomb} suggest that the on-site Hubbard 
repulsion $U$ is 
the largest energy in the problem. Its effects are typically considered
within the one- or two-orbital model whereby the strongly-correlated behaviour 
is induced by an interaction between the two spin-majority electronic $e_g$ 
states on-site, or between the two spin-components of a single $e_g$ 
band\cite{prb05}.

\begin{figure}
\includegraphics{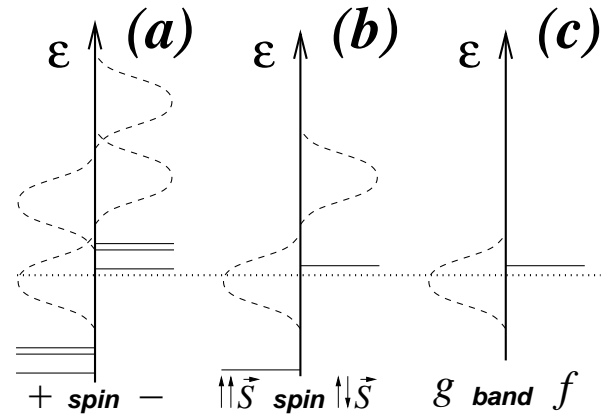}
\caption{\label{fig:scheme} (a) Crystal field splits the d-electron 
states of a Mn ion into $t_{2g}$ (solid) and $e_g$ (dashed); these split 
further due to a lattice distortion. Spin degeneracy is broken
due to Hund's rule, and the ``spin-plus'' $t_{2g}$ electrons form
the local spin 3/2. Chemical potential (dotted line) lies within an $e_g$ 
band, with a ``spin-minus'' $t_{2g}$ level nearby. 
(b) A simplified model of Eq.(\ref{eq:Ham0}),
with the spin quantisation axis fixed by a local spin $S$. (c) 
Relevant orbitals in the limit $J_H\rightarrow \infty$ (see text); adding an 
itinerant (localised) electron increases (reduces) the total on-site spin
of $S+1/2$ by $1/2$.}  
\end{figure}

It should be noted that the experimental data\cite{Nadgorny} indicate
the presence of {\it spin-minority} electrons near the Fermi level even
in the low-temperature ferromagnetic state. This agrees with the bandstructure
calculations \cite{bandcoulomb,bandminor}, suggesting that a narrow 
spin-minority band lies close to
the Fermi energy. Since both the localised $t_{2g}$ and itinerant $e_g$ 
states originate from
the same $d$-shell of a Mn ion, and therefore
are characterised by approximately the same value of Hund's rule splitting 
$J_H$, 
it is clear (see Fig. \ref{fig:scheme} {\it a}) 
that these spin-minority electrons populate the spin-down $t_{2g}$ (localised)
states\cite{bandcoulomb}.  This is further corroborated by the studies of 
${\rm La_{1-x} Ce_{x} Mn O_3}$ 
(with $1+x$ conduction electrons per formula unit) \cite{cerium}, which show 
both the 
spin-minority character of the carriers and a large overall increase
in the resistivity (in comparison with the usual $1-x$-electron case, 
{\it e.g.}, ${\rm La_{1-x} Ca_{x} Mn O_3}$), 
consistent with the extra $x$ electrons going into the spin-down 
$t_{2g}$ states. Yet, while ``two-fluid'' models involving 
itinerant and localised states were suggested earlier by both 
experimentalists and 
theorists\cite{Salamon,epl08,Rama},
the appealing possibility (mentioned in Ref. \cite{Chun95}) 
of these states having antialigned spins 
has not been addressed theoretically. 

Furthermore, we note that  electronic 
properties 
of a model where the orbital degree of freedom   
is taken into account are strongly coupled to lattice dynamics 
via the Jahn-Teller
effect \cite{Tokurabook}, which results in additional splitting of both
$e_g$ and $t_{2g}$ levels. However, the phenomenon of CMR occurs in a broad 
class of bulk systems [three-dimensional (3D) perovskites and quasi 
two-dimensional (quasi-2D) bilayered] and thin films of varying chemical 
composition, and is presumably always due to the same
physical mechanism. It is therefore worthwhile to consider a minimal model 
with fewer orbitals, which still captures some of the important 
{\it intra-atomic} physics of a Mn ion, before pursuing  more
complicated (and probably more 
material-specific) options\cite{single}. 

In the present Letter, we introduce such a simplified description and proceed 
with a simple mean-field analysis. While finer theoretical tools
are required to gain a fuller picture, qualitatively our results for 
electron dispersion and magnetic properties appear very encouraging.  

We consider a model involving 
a large spin $\vec{S}$ and two conduction-electron 
orbitals (broadened and nearly-localised) at each Mn site
(cf. Fig.  \ref{fig:scheme} {\it b}):
\begin{eqnarray}
\!\!\!\!\!&&\!\!\!\!\!{\cal H}=-\frac{t}{2} \sum_{\langle i,j \rangle,\alpha} 
\left(c^\dagger_{i \alpha}c_{j \alpha} +c^\dagger_{j \alpha}c_{i
\alpha}\right) + E_d^{(0)} \sum_{i,\alpha} 
d^\dagger_{i \alpha} d_{i \alpha}-  \nonumber \\
&&\!\!\!\!\!- \frac{J_H}{S} \sum_{i} \vec{S}_i \vec{\sigma_i}+
\frac{J}{S^2}\sum_{\langle i, j \rangle} \vec{S}_i \vec{S}_j -
\frac{H}{S}\sum_i\left( S^z_i+\sigma^z_i \right)
+\nonumber \\
&&\!\!\!\!\!+U \sum_i \left(c^\dagger_{i \uparrow} c^\dagger_{i \downarrow}
c_{i \downarrow}  c_{i \uparrow} + d^\dagger_{i \uparrow} 
d^\dagger_{i \downarrow}
d_{i \downarrow}  d_{i \uparrow}+\sum_{\alpha,\beta} c^\dagger_{i \alpha} 
d^\dagger_{i \beta}
d_{i \beta} c_{i \alpha} \right)- \nonumber \\
&&\!\!\!\!\!-\frac{V}{2}\sum_{\langle i,j \rangle,\alpha} 
\left(c^\dagger_{i \alpha}d_{j \alpha} +c^\dagger_{j \alpha}d_{i
\alpha}+ d^\dagger_{i \alpha}c_{j \alpha} +d^\dagger_{j \alpha}c_{i
\alpha} \right). \label{eq:Ham0}
\end{eqnarray}

Here, the operators $c_{i \alpha}$ ($d_{i\alpha}$) annihilate an $e_g$ 
($t_{2g}$) conduction electron of spin $\alpha = \uparrow,\downarrow$ (in
the laboratory frame) 
at a site $i$ of a square (or simple cubic) lattice.   
Localised spins $S_i$ originate from the remaining two $t_{2g}$ electrons,
hence, in reality $S=1$. They interact via superexchange $J$ and are also 
coupled to the spins of conduction electrons on-site, $\vec{\sigma}_i=
\frac{1}{2}\sum_{\alpha, \beta}\vec{\sigma}_{\alpha \beta}
(c^\dagger_{i \alpha}c_{i \beta}+d^\dagger_{i \alpha} d_{i \beta})$
(where $\vec{\sigma}_{\alpha \beta}$ are the Pauli matrices) via a strong 
ferromagnetic
Hund's rule exchange $J_H$; the external magnetic field $H$ is applied 
along the $z$-axis.
Owing to the fact that the (electron) co-ordinate operator
is not diagonal in the band index, 
there is a hybridisation $V$ between the
$t_{2g}$ and $e_g$ states; 
$E_d^{(0)}$ is the (bare) energy of the $t_{2g}$ electrons.
Direct hopping 
between $t_{2g}$ states on
different sites is assumed to be negligible, while the nearest-neighbour
hopping $t$ between the $e_g$ states will be used as an energy unit, $t=1$.

We construct the spin-wave expansion for the model (\ref{eq:Ham0}), keeping
terms up to first order in $1/S$.
In the spirit of Ref. \cite{Chubukov}, it is expedient to introduce a new
basis of electron states on each site according to
\begin{eqnarray*}
c_{\uparrow} \approx g_{\uparrow}-\frac{1}{\sqrt{2S}}g_{\downarrow}
\beta^\dagger-\frac{1}{4S}\left(g_{\uparrow} \beta^\dagger \beta + 
g^\dagger_{\downarrow} g_{\downarrow} g_{ \uparrow} + 
f^\dagger_{\downarrow} g_{\downarrow} f_{\uparrow} \right),\\
c_{\downarrow} \approx g_{\downarrow}+\frac{1}{\sqrt{2S}}g_{\uparrow}
\beta-\frac{1}{4S}\left(g_\downarrow+g_{\downarrow} \beta^\dagger \beta -
g^\dagger_{\uparrow} g_{\uparrow} g_{ \downarrow} - 
f^\dagger_{\uparrow} g_{\uparrow} f_{\downarrow} \right)
\end{eqnarray*}
(expressions for $d_\uparrow$ and $d_\downarrow$ are obtained by substituting
$g_\alpha \leftrightarrow f_\alpha$).
Operators $g_{i \uparrow}$ and $f_{i\uparrow}$ ($g_{i \downarrow}$ and 
$f_{i\downarrow}$) correspond, respectively, to the $e_g$ and $t_{2g}$ electrons
with a spin parallel (antiparallel) to the total spin on-site, 
$\vec{{\cal T}_i}=\vec{S_i}+\vec{\sigma_i}$, whose vibrations are annihilated
by a Holstein--Primakoff magnon operator $\beta_i$.
The Hund's rule term takes form
\begin{eqnarray*}
&&-\frac{J_H}{S}\vec{S}_i \vec{\sigma}_i=
-\frac{J_H}{2}\left\{g^\dagger_{i\uparrow}
g_{i \uparrow}+  f^\dagger_{i\uparrow} f_{i \uparrow}- (1+\frac{1}{S})
(g^\dagger_{i\downarrow}
g_{i \downarrow}+ \right. \\
&&\left.+f^\dagger_{i\downarrow} f_{i \downarrow})   
+\frac{1}{S} \left[g^\dagger_{i\uparrow}g^\dagger_{i\downarrow}
g_{i\downarrow} g_{i\uparrow}-g^\dagger_{i\uparrow}f^\dagger_{i\downarrow}
g_{i\downarrow} f_{i\uparrow}+(f \leftrightarrow g)\right]\right\}\,.
\end{eqnarray*}

Finally, the Holstein--Primakoff operators $a_i$ of the original spins 
$\vec{S}_i$ are expressed as
\[a \approx \beta-\frac{g^\dagger_\uparrow g_\downarrow  
+f^\dagger_\uparrow f_\downarrow}{\sqrt{2S}} -\frac{g^\dagger_\uparrow
g_\uparrow+f^\dagger_\uparrow f_\uparrow-g^\dagger_\downarrow g_\downarrow
-f_\downarrow^\dagger f_\downarrow }{4S} \beta. \]

We next substitute these expressions into the Hamiltonian (\ref{eq:Ham0})
and take the limit of $J_H \rightarrow \infty$ while keeping 
$E_d=E_{d}^{(0)}+J_H$ constant. Now, if the chemical potential,
denoted $\mu +U - J_H/2$, lies within the spin-up $e_g$ band, then the
spin-down $e_g$ band is completely empty, and the spin-up $t_{2g}$ band
completely filled. This is precisely the case of interest to us, containing
the effects of the Coulomb repulsion between the spin-up $e_g$ and spin-down
$t_{2g}$ electrons in the presence of a localised spin $S+1/2$ 
(Fig. \ref{fig:scheme} {\it c}). Henceforth,
we drop all the terms containing the operators $g_\downarrow$ and $f_\uparrow$,
and suppress the spin index of remaining fermion operators.  

Assuming the ferromagnetic ground state, the 
Hamiltonian takes the form
${\cal H}_e + {\cal H}_h + {\cal H}_m$ with the electronic and 
(magnon-assisted) hybridisation terms,
\begin{eqnarray}
\!\!\!\!\!{\cal H}_e&\!\!\!\!\!\!=&\!\!\!\!\!\!\sum_{\vec{k}} \left(\epsilon_{\vec{k}}- \mu \right) g^\dagger_{\vec{k}} 
g_{\vec{k}}+ \left(E_d- \mu \right) \sum_{j} f^\dagger_{j} 
f_{j}+ \nonumber \\
&+&\frac{U}{N} {\sum_{\vec{k}, \vec{k}^\prime,j}} 
{\rm e}^{i(\vec{k}^\prime -\vec{k})\vec{R}_j} g^\dagger_{\vec{k}} 
f^\dagger_j f_j
g_{\vec{k}^\prime} \,, \label{eq:hefm}\\  
\!\!\!\!\!{\cal H}_h&\!\!\!\!\!\!=&\!\!\!\!\!\!-\frac{V}{\sqrt{2S}N}\sum_{\vec{k}, \vec{q},j} 
(\epsilon_{\vec{k}}\!-\!\!\epsilon_{\vec{k}+\vec{q}}) 
{\rm e}^{-i(\vec{k} +\vec{q})\vec{R}_j}
g^\dagger_{\vec{k}} f_j
 \beta^\dagger_{\vec{q}}\!+\! {\rm H. c.} \label{eq:hhfm}
\end{eqnarray}
Here, $N$ is the number of lattice sites, $\epsilon_{\vec k}=-\cos k_x
-\cos k_y (-\cos k_z)$ is the tight-binding dispersion law in two (three)
dimensions, and ${\vec R}_j$ is the radius-vector of site $j$. It is assumed
that the site basis is more appropriate for describing the narrow-band
fermions $f_j$, easily localised by fluctuations or disorder.
The term ${\cal H}_m$, which is of order $1/S$, contains  Zeeman electron energy shifts, magnon dynamics, and
double-exchange band-narrowing effects:
\begin{eqnarray}
&&{\cal H}_m=\frac{H}{2S}\!\left(\sum_j f^\dagger_j f_j \!\!-\sum_{\vec{k}}g^\dagger_{\vec{k}}
g_{\vec{k}}\right) \!\!+\nonumber \\&&+ \frac{1}{S}\sum_{\vec{k}}\left[H-2J(\epsilon_{\vec{k}}+d)\right]\beta^\dagger_{\vec{k}} \beta_{\vec{k}}+ \nonumber \\
&&+\frac{1}{4SN} {\sum_{1 \div 4}}^\prime\left(2\epsilon_{2+4}-
\epsilon_1-\epsilon_2\right) g^\dagger_1 g_2 \beta^\dagger_3 \beta_4\,. \label{eq:hmfm}
\end{eqnarray}
Here, momentum-conserving summation is denoted by $\Sigma^\prime$,
and $d$ is the dimensionality of the system (2 or 3).

The electronic term, ${\cal H}_e$, is the familiar Falikov--Kimball model.
The rich physics contained therein\cite{FKrefs} crucially
depends on the presence (and nature) of the inter-band hybridisation.
The form of our ${\cal H}_h$, Eq. (\ref{eq:hhfm}), is dictated by spin
conservation: electron transfer between the two opposite-spin bands
must be accompanied by magnon creation or annihilation, 
$\beta^\dagger_{\vec{q}}$ or
$\beta_{\vec{q}}$. Such transfers require misalignment of spins
$\vec{S}_i$ on {\it neighbouring} sites, hence the
hybridisation matrix element vanishes at $q \rightarrow 0$, underlining
the importance of short-wavelength processes. The latter feature
appears promising in the context of CMR compounds, where the 
 unusual short-range correlations are reflected in the electronic and 
magnetic\cite{Lynn,Lynnreview} properties. Importantly, magnetic 
field $H$ affects the carriers both via double exchange mechanism and by 
changing the 
energy difference between localised and itinerant states 
[see Eq. (\ref{eq:hmfm})].

We shall be interested in the regime characterised by 
non-zero values of {\it both} fermion occupancies $n^g$ and  $n^f$
(the latter assumed independent on $\vec{R}_i$),
\begin{equation}
n^f=\langle f^\dagger_i f_i\rangle,\,\,\,\,
n^g=\frac{1}{N}\sum_{\vec{k}}n^g_{\vec{k}}\,,\,\,\,\,
n^g_{\vec{k}}=\langle g^\dagger_{\vec{k}} g_{\vec{k}} \rangle.
\label{eq:diag}
\end{equation}
In addition, there also arises an off-diagonal
average,
\begin{equation}
\langle 
f^\dagger_j g_{\vec{k}} \beta_{\vec{q}}
\rangle \equiv \langle 
f^\dagger g_{\vec{k}} \beta_{\vec{q}}
\rangle {\rm e}^{-i(\vec{k}+\vec{q})\vec{R}_j}  
\label{eq:offdiag}
\end{equation}
(here and below, we omit the site indexes of operators $f_j$ once the
$j$-dependent exponent has been factored out). 

In order to clarify the basic physics contained in our model, Eqs. 
(\ref{eq:hefm}--\ref{eq:hmfm}), we 
will now proceed with a mean-field analysis of it. 
Here, we focus on the simplest self-consistent scheme, allowing only
for average values (\ref{eq:diag}-\ref{eq:offdiag}) and for a non-zero 
magnon occupancy. While actual validity of this approach is probably
restricted to the intermediate temperature range  (on the scale
of the Curie temperature) and moderate
values of $U$ (see below), it offers important guidance for future 
investigations.
Mean field equations can be found in a standard way by decoupling the
equations of motion for the appropriate retarded Green's functions, expressing
the latter as

\begin{widetext}

\begin{eqnarray}
\llangle f f^\dagger \rrangle &\equiv& \llangle f_j f^\dagger_j \rrangle=
\left\{\omega -\tilde{E}_d+\mu+i0 -\frac{V}{2SN^2}\sum_{\vec{k},\vec{q}}
\frac{V(1+{\cal}N_{\vec{q}}-n^g_{\vec{k}})(\epsilon_{\vec{k}}-
\epsilon_{\vec{k}+\vec{q}})^2 +U \Phi_{\vec{q}}(\epsilon_{\vec{k}}-
\epsilon_{\vec{k}+\vec{q}})}{\omega+i0-\tilde{\epsilon}_{\vec{k}}+\mu}\right\}^{-1}\,, \label{eq:Gff} \\
\llangle g_{\vec{k}}g^\dagger_{\vec{k}}\rrangle&=& \left\{\omega -
\tilde{\epsilon}_{\vec{k}}
+\mu+i0 -\frac{V}{2SN}
\frac{\sum_{\vec{q}}\left[V(N_{\vec{q}}+n^f)(\epsilon_{\vec{k}}-
\epsilon_{\vec{k}+\vec{q}})^2 +U \Phi_{\vec{q}}^*(\epsilon_{\vec{k}}-
\epsilon_{\vec{k}+\vec{q}})\right]}{\omega+i0-\tilde{E}_{d}+\mu}\right\}^{-1}\,,\label{eq:Ggg}\\
\llangle \beta_{\vec{q}}\beta^\dagger_{\vec{q}}\rrangle&=&
\left\{\omega -\omega^0_{\vec{q}}+i0+\frac{V^2}{2NS}\sum_{\vec{k}} 
\frac{(n^f-n^g_{\vec{k}})(\epsilon_{\vec{k}}-
\epsilon_{\vec{k}+\vec{q}})^2}{\omega+i0-\tilde{E}_d+\tilde{\epsilon}_{\vec{k}}}+\frac{V^2U}{2SN^2}\frac{\left[\sum_{\vec{k}} 
\frac{(n^f-n^g_{\vec{k}})(\epsilon_{\vec{k}}-
\epsilon_{\vec{k}+\vec{q}})}
{\omega+i0-\tilde{E}_d+\tilde{\epsilon}_{\vec{k}}}\right]^2}{1-
\frac{U}{N}\sum_{\vec{k}} 
\frac{n^f-n^g_{\vec{k}}}
{\omega+i0-\tilde{E}_d+\tilde{\epsilon}_{\vec{k}}}}
\right\}^{-1}\,,  \label{eq:Greenmagnon}\\
\Phi_{\vec{q}}&=&\Phi^*_{\vec{q}}\equiv \sqrt{2S}\sum_{\vec{k}}\langle 
g^\dagger_{\vec{k}}f \beta^\dagger_{\vec{q}}\rangle=-\frac{V}{\pi}\int
{\rm Im} \left\{ \frac{\frac{1}{N}\sum_{\vec{k}} 
\frac{(n^f-n^g_{\vec{k}})(\epsilon_{\vec{k}}-
\epsilon_{\vec{k}+\vec{q}})}
{\omega+i0-\tilde{E}_d+\tilde{\epsilon}_{\vec{k}}}}{1-
\frac{U}{N}\sum_{\vec{k}} 
\frac{n^f-n^g_{\vec{k}}}
{\omega+i0-\tilde{E}_d+\tilde{\epsilon}_{\vec{k}}}} \llangle \beta_{\vec{q}}\beta^\dagger_{\vec{q}}\rrangle \right\} \frac{d \omega}{\exp(\frac{\omega}{T})-1}\,. \label{eq:Phi}
\end{eqnarray}
\end{widetext}
Here, the magnon occupancy is 
\begin{equation}
{\cal N}_{\vec{q}} \equiv \langle \beta^\dagger_{\vec{q}} \beta_{\vec{q}}
\rangle
= -\frac{1}{\pi} \int {\rm Im} 
\llangle \beta_{\vec{q}}\beta^\dagger_{\vec{q}}\rrangle \frac{d \omega}
{\exp(\omega/T)-1}\,
\end{equation}
($n^g_{\vec{k}}$ and $n^f$ are expressed in a similar way). 
Hartree energies of magnons and 
those of $e_g$ and $t_{2g}$ electrons read
\begin{eqnarray}
\!\!\!\omega^0_{\vec{q}}&=&\frac{H}{S}-\frac{2J}{S}(\epsilon_{\vec{q}}+d)+
\frac{1}{2NS}\sum_{\vec{k}}n^g_{\vec{k}}(\epsilon_{\vec{k}+\vec{q}}-
\epsilon_{\vec{k}}), \label{eq:omega0}\\
\!\!\!\tilde{\epsilon}_{\vec{k}}&=&\epsilon_{\vec{k}}+Un^f-\frac{H}{2S}+
\frac{1}{2NS}\sum_{\vec{q}}{\cal N}_{\vec{q}}(\epsilon_{\vec{k}+\vec{q}}-
\epsilon_{\vec{k}}),\label{eq:epsilont}\\
\!\!\!\tilde{E}_d&=&E_d+\frac{U}{N}\sum_{\vec{k}}n^g_{\vec{k}}+\frac{H}{2S},
\end{eqnarray}
The last terms in 
Eqs. (\ref{eq:omega0}-\ref{eq:epsilont}) contain the familiar double exchange
physics (ferromagnetic contribution to the spin-wave spectrum and the 
magnon-induced conduction band narrowing, respectively). 

\begin{figure*}
\includegraphics{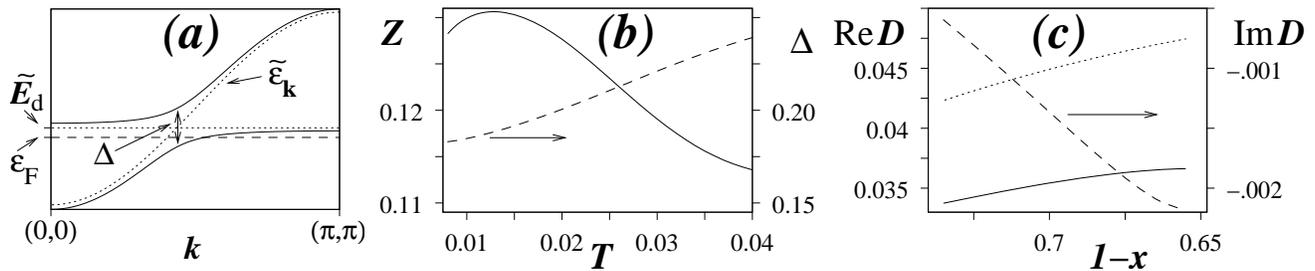}
\caption{\label{fig:fermion}Mean field results for a 2D system with
$E_d=-0.35$, $V=0.2$, $U=1.2$, and $J=0$. {\it (a)} Schematic view of the 
itinerant fermion dispersion (solid lines); dotted lines show the 
unhybridised $\tilde{\epsilon}_k$ and $\tilde{E}_d$. {\it (b)}:
Temperature dependence of quasiparticle weight $Z$ at 
the Fermi level (solid line) and the gap, $\Delta(T)$ (dashed line), 
for the electron
density $1-x=0.7$. {\it (c)} Doping dependence of real (solid line) and 
imaginary (dashed) parts of spin stiffness $D$ at $T=0.0015$. Dotted line 
shows spin 
stiffness for a usual double exchange model. $t_{2g}$ occupancy increases
from $n_f=0.25$ at $1-x=0.65$ to $n_f=0.43$ at $1-x=0.73$.}
\end{figure*}

In analysing Eqs. (\ref{eq:Gff}--\ref{eq:Phi}) we consider a 2D  
system; the 3D case can be expected to be similar.
We replace 
all factors $(\epsilon_k-\epsilon_{k-q})^2$ with their average 
values over the isoenergetic surfaces\cite{replace} 
$\epsilon=\epsilon_{\vec{k}}$ and $\epsilon=\epsilon_{\vec{q}}$. We 
arrive at a system of mean field 
equations for $\mu$, $n_b$, and four quantities $\int\! \Phi(\epsilon) (2+\epsilon) \nu d \epsilon$,$\int\! {\cal N}(\epsilon)
\epsilon\nu d\epsilon$, 
$\int\! {\cal N}(\epsilon)\langle v^2 \rangle_\epsilon \nu d\epsilon$,
and $\int\!{\cal N}(\epsilon)(4-\epsilon^2) \nu d\epsilon$,
where $\nu(\epsilon)$ and  $\langle v^2 \rangle_\epsilon$ are the tight binding
density of states and average velocity square at a given energy $\epsilon$.
Solving these equations numerically, we observe  that:

\noindent
(i) The localised band is broadened, and a temperature-dependent gap $\Delta$
(a new small {\it energy scale}) opens 
in the spectrum of itinerant electrons (fermions $g_{\vec{k}}$). Quasiparticle
weight of itinerant electrons decreases when the energy approaches the gap 
from either 
side. The Fermi level lies below the gap, and the {\it quasiparticle
weight at the Fermi surface is strongly suppressed} (Fig. \ref{fig:fermion} 
{\it a} and {\it b}). 
This behaviour,
which is already reminiscent of a T-dependent (pseudo)-gap found 
experimentally\cite{pseudogap},
will be further modified in a more exact treatment ({\it e.g.}, a finite 
relaxation time will arise in the second order in 1/S). Eqs. (\ref{eq:Gff}--
\ref{eq:Ggg}) imply that in the present model, these spectral features
are directly controlled by spin dynamics.

\noindent
(ii) With only the spin-majority electrons contributing to the spin stiffness
$D$, 
which in 2D or 3D is given by
\[
DS=-\frac{1}{4dN}\sum_{\vec{k}} \epsilon_{\vec{k}}n^g_{\vec{k}}-J-\frac{V^2}{2dN}\sum_{\vec{k}} \frac{n^f-n^g_{\vec{k}}}
{\tilde{\epsilon}_{\vec{k}}-\tilde{E}_d+i0} \left(\frac{\partial \epsilon_{\vec{k}}}
{\partial \vec{k}}\right)^2  \,,
\]
the latter 
is suppressed in comparison with the usual double exchange case 
(Fig. \ref{fig:fermion} {\it c}).
An unusual feature of the present model is that the spin-flip continuum 
extends down to zero energy and momentum\cite{nopole}. As a result,  $D$ also 
develops  an imaginary part.
This implies {\it strong magnon damping},  as observed experimentally
\cite{Lynnreview}. Magnon damping proportional to $q^2$
is not usually expected in a ferromagnet\cite{Lynnreview,Tserk} 
and means that the spin-diffusion coefficient
acquires a real (dissipative) part.

As for the diffusive central peak found in the inelastic neutron 
scattering\cite{Lynn}, we expect it to arise once the magnon-assisted 
diffusive motion of $t_{2g}$ electrons (neglected here) is taken into account. 
Experimentally, strongly damped magnons,
central peak, and pseudogap in the density of 
states\cite{pseudogap} (or optical Drude weight reduction\cite{Okimoto})
are the key 
generic features of CMR manganates at the intermediate-to-high temperatures 
below $T_C$. 

A relatively small value of $U$ used in Fig. \ref{fig:fermion}
is due to the reduced stability 
region for mean field solutions with both $n_f$ and $n_g$ different from zero. 
This
reduction is an expected artifact of a simplistic mean field approach, 
mirroring, {\it e.g.}, the greatly enhanced mean field stability of 
ferromagnetism in the 
Hubbard model. This situation calls for further theoretical investigation, 
combining advanced mean-field schemes with numerical methods.

These future treatments will also have to address the issue of ferro- to 
paramagnetic transition and a possibility of charge ordering. We expect
that any transition will be accompanied by a change of electron distribution
between the two bands, thus changing the magnitude of the net spin ${\cal
T}_i$ 
on-site. Experimentally, the relevant quantity
is the average total spin $\langle {\cal T} \rangle=(3+n^g-n^f)/2$ of a Mn ion,  
which should show temperature and magnetic field dependence, especially 
in the region around and above $T_C$. In particular, this should lead
to an unconventional
{\it longitudinal} spin dynamics\cite{Atsarkin} and to a renormalisation
of the  Curie--Weiss
constant (cf. Ref. \cite{cwtemp}).  
In principle, the value of $\langle {\cal T} \rangle$ should also be 
accessible more directly via muon spin rotation \cite{muon} and 
NMR \cite{nmr} measurements.
We suggest that these methods (combined with  
electron spectroscopy) should be used to measure the value of 
$\langle {\cal T} \rangle$. Its temperature dependence, especially if it 
correlates
 with (magneto)transport properties, would
imply that a successful theoretical description of CMR compounds
should indeed include spin-minority localised electrons.

It is a pleasure to thank R. Berkovits, G. Goobes, A. Kanigel, 
K. A. Kikoin, E. M. Kogan, B. D. Laikhtman, D. Orgad, and S. Satpathy 
for discussions.
This work was supported by the Israeli Absorption Ministry.


\begin{thebibliography}{999}
\bibitem{replace} All the relevant quantities then depend on $\vec{k}$
via $\epsilon_{\vec{k}}$; this is not expected to affect the results.
\bibitem{nopole} Magnon Green's function, Eq. (\ref{eq:Greenmagnon}), has a 
branch-cut, but not necessarily
a pole. Magnon spectrum is defined as the location of spectral weight peak.
\end{thebibliography}
\end{document}